# Localized Load Reduction Market Development Considering Network Constraints

Mohammad Panahazari[1], Minoo Mohebbifar[2], Vahid Nazari Farsani[1], Mahmoud-Reza Haghifam[1]
[1] Department of Electrical and Computer Engineering, Tarbiat Modares, University, Tehran, Iran
[2] Department of Electrical and Computer Engineering, Tabriz, University, Tabriz, Iran

*Emails: {m.panahazari, nazarivahid, haghifam}@modares.ac.ir, mohebbifar.minoo@tabriz.ac.ir*

*Abstract*— With the development of the smart grid concept and the increasing expansion of advanced communication and measurement equipment, consumers can actively participate in the power system operation. The intelligent use of these facilities greatly helps the power system entities to achieve their objectives more efficiently and less expensively. As a beneficial facility, the market mechanism has proven to be a solution to various power system challenges. Furthermore, distributed and localized solutions have shown to be helpful in both reducing operation costs and accelerating the execution of the programs. In a generation shortage condition, to prevent unwanted load curtailment and wholesale market price spikes, utilities can get consumers' help to reduce the load in return for payments. This paper proposes a localized load reduction market model in the distribution system, in which consumers bid for their participation rate at the corresponding prices. Then, a market optimization problem will be solved by considering the technical constraints of the network through the use of Genetic Algorithm (GA). The paper then shows that utilizing the proposed model reduces operation costs.

## I. Introduction

Nowadays, consumers can actively participate in secure and optimal operation programs using smart meters and communication devices. System operators in particular Distribution System Operator (DSO) can use consumers' capabilities to balance electrical supply and demand by reducing the demand during critical periods instead of increasing the power generation. Moreover, consumers can help DSO improve the system's voltage and frequency stability [1]. The development of this idea has led to the concept of Demand Response (DR) and its application in power systems. DR methods can be divided into Price-based and Incentive-based groups. Price-based methods are suitable for small and medium-sized consumers, and Incentive-based methods are more suitable for large and industrial consumers [2]. Incentive-based methods can also be divided into classical and market-based methods. Direct load control (DLC) is one of the classic methods of DR. In [3] a new method of load reduction during peak hours, using the Internet of Things (IoT) is proposed. Survey [4] presents a two-step distributed direct load control approach to control customers' electrical appliances. With the aim of optimal utilization of DR, [5] shows a method of direct and indirect load control for congestion management.

Another classic method in DR programs is the Interruptible/Curtailable (I/C) load. In this program, a specific part of the electric load or the total consumption is curtailed in return for an incentive reward. The consumer will be penalized in the situation of not being committed to the contract [6]. In [7] uncertainties of I/C implementation are modeled, and the uncertainty curve of each consumer's participation in the program is calculated using past data. Then a level of participation reliability for consumers is introduced. Authors in [8] have presented a mathematical model of demand response to investigate the effect of tariffs on customer consumption during peak times and off-peak consumption. The effect of changes in incentives and penalties on the implementation of I/C service on the Microgrid is studied in [9] considering different uncertainties. Emergency Demand Response Programs (EDRP)s can be classified as market-based DR Programs. A game-theoretic model is proposed in [10] to determine discriminative rebates and scarcity pricing of electricity, encouraging industrial users to reduce their power usage during times of scarcity. The model is shown to improve profits for participants, with a higher rebate suggested for high-value-added and non-interruptible manufacturing. Another market-based demand response program is Demand Bidding, which is mainly for large-scale and industrial loads. In [11], a regulated peer-to-peer market structure for residential prosumers is proposed to regulate dynamic electricity prices in transactive energy systems. The proposed market structure enables prosumers to bid energy demand or supply curves directly into a peer-to-peer market, which can be used by the utility to quantify and dispatch demand flexibility. Also, it can be used to facilitate the conventional unit commitment problem. Moreover, other novel ways to design Incentive-based DR are presented in the articles [12]. Authors in [13] use the present Stackelberg game between utilities and end-users to maximize utility revenue and users' pay-offs. According to [14], in a multi-objective modeling approach, consumers bid for their participation in the market under an incentive-based DR program that considers both the economic drivers and social factors affecting the market actors, with a focus on customer satisfaction. Some articles have considered network technical constraints. The DR Program which is studied in most of them has been Direct Load Control. In [15], a DR Program is proposed to control the load of residential customers, and the planning of customers' consumption is done with the consideration of voltage and loss constraints. [16] Has presented a method of residential DLC for an unbalanced network trying to meet technical limitations of the network. One of the challenges of designing an efficient incentive-based program is to consider the uncertainties imposed by the daily wholesale market price, renewable sources' output power oscillation, and load changes. Therefore, some authors, inspired by the applications of artificial intelligence in the fields of power systems, such as load forecasting, energy theft detection [17], machine control [18], and motor fault detection [19], have proposed a combination of deep learning and recurrent neural networks to solve the above-mentioned uncertainty [20].
Most previous research on market-based DR has focused on the wholesale market, and few have talked about local markets. In addition to this, technical constraints have also been paid less attention. Consequently, this paper proposes a Localized Load Reduction Market (LLRM) in which consumers provide bids for their load reduction like a virtual power plant. Besides, the utility and other entities that need load reduction are buyers in this market. Furthermore, the paper focuses on the technical constraints of the network. This paper has benefited from GA to find the optimal plan of load reduction considering the constraints mentioned above. In the second part of the article, the proposed market is described. Section III discusses the proposed market optimization method which is GA. Later in section IV, the paper uses three scenarios to verify the model presenting details of case studies and simulation results. Finally, section V concludes the paper.

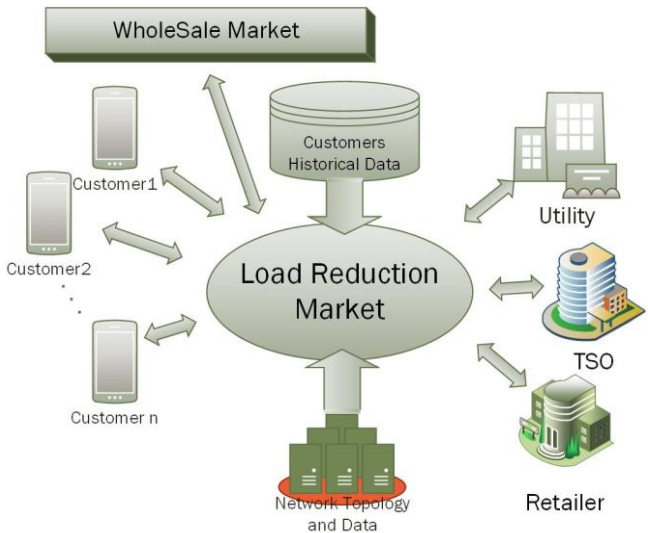

Fig. 1. Localized Load Reduction Market Framework

## II. MARKET DESCRIPTION

### A. Market Objecive

DR Programs are used to encourage customers to decrease the load in high wholesale electricity prices, or reliability-triggered events [21]. Considering a high-demand circumstance under which the network has faced congestion in a feeder, it is clear that, the operator has to curtail a part of the load to deal with the problem. The existing LLRM helps utilities to use consumers' participation in the market to reduce as much load as they need having the least possible cost. The objective of LLRM can vary depending on the entity that executes it.

### B. Choosing Market Runner

As mentioned earlier, the market's objectives can vary depending on the entity that runs it. In this paper, the utility is the runner of LLRM. To investigate the purpose of the utility for running LLRM, we need a precise definition of the utility. The utility duties and benefits are different depending on different management and control structures. Generally, utilities can play three roles in the system: 1) Wire Company and Retailer; 2) Wire Company; 3) Retailer. In this paper, the power system is considered to be totally restructured, and retailers have a role in the system. As a result, the utility role is just Wire Company.

Similarly, there is also a situation in which the utility has just a Wire Company role, and that is when there are forced tariffs for electricity price. In this situation, the more power the utility can dispatch the more income it will have. On the other hand, the regulator penalizes the utility in curtailments.

### C. Market Framework Description

As shown in Fig.1 the key parts of LLRM are the market runner entity and its objective, the customers that bid for the reduction, the network topology and data, and importantly the historical data of customers. Different entities have different aims for load reduction. So, the entities can use LLRM to reach the objective with the least possible cost. LLRM framework is based on an application for registration of players, showing winners, giving archives and an environment for sending and receiving messages and requests. The market is initiated when an entity sends a request inviting consumers to participate, and registered consumers receive this message. Then, consumers start bidding for their load reduction like what a power plant does in the wholesale market. Market Operator (MO) collects the bids and checks the possibility of consumers' proposed reductions using archived historical data. After that, considering the runner entity and what objective the market has, MO clears the market and introduces the winners. At this step, the winners have to reduce the proposed load. Otherwise, they will be penalized.

## III. PROBLEM FORMULATION

Suppose that feeder $j$ faces congestion, and the utility wants to deal with the problem by decreasing consumption. So, the utility sends its request to LLRM, and the market invites consumers to participate in load reduction program. The utility tries to reduce all needed load using consumers' reductions. The other choice is load curtailment. By adopting this approach, the utility would be required to pay a predetermined penalty to compensate consumers for any lost load.

Demand in this paper is supposed to be divided into two parts: firmed and non-firmed part [22]. So, the utility can only curtail non-firmed part of the load. In the formulations, the maximum allowed demand to be curtailed named as non-firmed capacity for the consumer is $P_i^{max}$.

Consumers bid for the reduction using a step function $f_n(P_i)$ named as a function of power level that

$$p \in \{0, P_1, P_2, \ldots, P_i, \ldots, P^{max}\}.$$

Let us assume that consumer $i$ has suggested a price for reduction and the curtailment cost of that consumer for is $C_{cr}^i$. On the other hand, utilities are responsible for payment of network loss cost. Hence, the problem could be mathematically expressed as:

$$\min \left\{ \sum_{i=1}^{N} P_{cr}^i * C_{cr}^i + \sum_{i=1}^{M} P_{DR}^i * C_{DR}^i + C_{loss} \right\} \quad (1)$$

| **Algorithm1** Market clearing using Genetic Algorithm |
|---|
| **Step1.** Get network topology data, wholesale market price, forecasted demand, loads, consumers' load curtailment cost data and consumers' bids data. |
| **Step2.** Get Iteration count, mutation and crossover percentage. |
| **Step3.** Generate random feasible population. |
| **Step4.** Mutate determined percentage of population |
| **Step5.** Crossover determined percentage of population. |
| **Step6.** Evaluate fitness of population and select best solution. |
| **Step7.** If not reached maximum number of iteration go to step4. |

The constraints are as follows:

$$P_{cr}^i < P_{cr}^{i\,max} \quad (2)$$

$$P_{DR}^i < P_{DR}^{i\,max} \quad (3)$$

$$P_{DR}^i < P_{cr}^{i\,max} - P_{cr}^i \quad (4)$$

$$\{\sum_{i=1}^{N} P_{cr}^i + \sum_{i=1}^{M} P_{DR}^i\} > P_{sch} \quad (5)$$

$$I_l < I_l^{max} \quad \forall\, l \in j \quad (6)$$

$$V_l > V_l^{min} \quad \forall\, l \in j \quad (7)$$

Where:

$P_{cr}^{i\,max}$ : maximum curtailable load for i-th consumer;

$P_{DR}^{i\,max}$ : maximum suggested reduction by consumer;

$P_{sch}$ : total scheduled load for reduction;

$I_l$ : current of l-th branch;

$I_l^{max}$ : maximum current of l-th branch;

$V_l$ : voltage of l-th bus;

$V_l^{max}$ : maximum voltage of l-th bus;

Eq. (2) – (7) represent the constraints of the optimization problem. The constraint in eq. (5) Is used when market runner wants to reduce consumption to a specific amount. Eq. (6) is the congestion limit and will be added to the model if the runner wants to resolve the congestion problem in the network. Also, Eq. (7) Is the voltage limit. To solve the optimization problem, we consider and as integer variables. We use GA in order to find the optimal point, and it is described in the next section.

## IV. OPTIMIZATION ALGORITHM

The market runner needs to select the best group of suggestions with an economic point of view considering network constraints. GA is a highly effective search algorithm that has been widely implemented in various research studies. [23]. For this reason, in this paper, GA is used for selecting presented suggestions by consumers. For clearing the market with GA, network information, including topology and consumers' consumption forecasts are collected for the desired time. Then, consumers' bids and their curtailment costs are entered. After that, an initial population is created from feasible answers considering consumers' bids and their curtailment costs. The feasibility of these answers varies depending on the market runner entity. For example, when the utility is the runner, intending to resolve the congestion in the network, the answers should not only meet load constraints but also relieve the congestion in the network. GA repeats for a predefined number of iterations. In each iteration, the algorithm chooses some of the individuals of the population. Then, after performing crossover and mutation operators, a truncation procedure should be used to guarantee that each gene takes an integer value. Then, old selected individuals will be replaced with newly created ones. The algorithm mentioned above is described in Algorithm 1.

## V. SIMULATION RESULTS

In order to illustrate the application of the proposed method, IEEE 33-node test feeder is considered. For this network total load without reduction of demand is 3715 kW and the loss is 201.9 kW. Considering [24], the cost of load curtailment is various not only for different consumers but also for similar ones from different countries. The other factors that affect load curtailment cost are duration, time of day, day of the week, etc. Table I contains samples of load curtailment costs used in this paper for simulation.

Suppose that consumers provide bids in 10 kW steps, and the maximum proposed demand by consumers for reduction is 130 kW. Table II includes consumers' bid steps for participation in LLMR based on the research that has been introduced in [21]. It is assumed that consumers are capable of calculating approximately the best price for their bids. So, the article has skipped the issue. In order to simulate the market, three scenarios have been chosen.

### A. Using market for voltage regulation

In this scenario, the utility aims to regulate network voltage with the utilization of LLMR. Let us assume that the maximum allowed dropped voltage is 5 percent and the wholesale electricity price is 40$/MWh. In addition to this, constraints (5) and (6) are neglected because the congestion is not essential for the utility in this scenario. Furthermore, the amount of reduced load does not matter in this situation. After execution of the program and clearing LLMR using GA, the desired result was obtained. As seen in Figure 2, the maximum dropped voltage has reached 5 percent, the total cost of the program has been $454.66, and feeder loss has been reduced to 79 kW. Additionally, Figure 3 shows the maximum suggested load by consumers and their curtailed load by the market. Without using LLMR, the cost of voltage regulation increases dramatically, reaching $41,328.

TABLE I. CONSUMERS' LOAD CURTAILMENT COST

| Consumer | Firmed Load (kW) | Non-Firmed Load (kW) | Curtailment Cost ($/kWh) |
|---|---|---|---|
| 1 | 0 | 0 | - |
| 2 | 80 | 20 | 73 |
| 3 | 70 | 20 | 43 |
| 4 | 100 | 20 | 55 |
| 5 | 40 | 20 | 37 |
| 6 | 30 | 30 | 80 |
| 7 | 100 | 100 | 59 |
| 8 | 60 | 140 | 45 |
| 9 | 50 | 10 | 60 |
| 10 | 50 | 10 | 70 |
| 11 | 40 | 5 | 90 |
| 12 | 40 | 20 | 55 |
| 13 | 20 | 40 | 60 |
| 14 | 100 | 20 | 60 |
| 15 | 50 | 10 | 45 |
| 16 | 50 | 10 | 80 |
| 17 | 40 | 20 | 85 |
| 18 | 60 | 30 | 65 |
| 19 | 30 | 60 | 60 |
| 20 | 20 | 70 | 68 |
| 21 | 70 | 20 | 54 |
| 22 | 60 | 30 | 40 |
| 23 | 60 | 30 | 90 |
| 24 | 200 | 220 | 55 |
| 25 | 250 | 170 | 61 |
| 26 | 30 | 30 | 55 |
| 27 | 10 | 50 | 60 |
| 28 | 40 | 20 | 80 |
| 29 | 100 | 20 | 60 |
| 30 | 150 | 50 | 55.5 |
| 31 | 30 | 120 | 50 |
| 32 | 140 | 70 | 70 |
| 33 | 20 | 40 | 70 |

### B. Using market for congestion management

In this scenario, the utility uses LLMR to relieve the congestion in the feeder. Maximum congestion limit for branches of the feeder is supposed to be 0.04 p.u. Fig.4 presents branch currents before and after LLRM execution. It shows that the congestion of the network has been completely relieved. The amount of curtailed load by the market for this scenario and the maximum proposed load by consumers is shown in Fig. 5. Total cost in this scenario has been 141.7$, and loss has reduced to 137.1 kW. If the utility decides to resolve the congestion without LLMR, it would cost 15560$ that is almost 100 times greater than the LLMR cost.

TABLE II. CONSUMERS' BIDS

| Consumer | 2 | 3 | 4 | 7 | 9 | 10 | 11 | 12 | 14 | 16 | 17 | 18 | 22 | 23 | 24 | 25 | 29 | 30 | 32 | 33 |
|---|---|---|---|---|---|---|---|---|---|---|---|---|---|---|---|---|---|---|---|---|
| Power(kW) | ($/kwh) Bids | | | | | | | | | | | | | | | | | | | |
| 10 | 0.3 | 0.22 | 0.29 | 0.33 | 0.4 | 0.45 | 0.4 | 0.28 | 0.25 | 0.32 | 0.5 | 0.24 | 0.23 | 0.5 | 0.32 | 0.19 | 0.35 | 0.25 | 0.26 | 0.4 |
| 20 | 0.35 | 0.26 | 0.31 | 0.35 | 0.5 | 0.45 | 0.4 | 0.42 | 0.3 | 0.35 | 0.6 | 0.26 | 0.27 | 0.6 | 0.34 | 0.24 | 0.37 | 0.27 | 0.29 | 0.6 |
| 30 | 0.4 | 0.3 | 0.33 | 0.37 | 0.6 | 0.45 | 0.4 | - | 0.35 | 0.38 | 0.7 | 0.28 | 0.31 | 0.7 | 0.36 | 0.29 | 0.39 | 0.29 | 0.32 | - |
| 40 | 0.45 | 0.34 | 0.35 | 0.39 | 0.7 | - | - | - | 0.4 | 0.41 | 0.8 | 0.3 | 0.35 | 0.8 | 0.38 | 0.34 | 0.41 | 0.31 | 0.35 | - |
| 50 | 0.5 | 0.38 | 0.37 | 0.41 | - | - | - | - | 0.45 | - | - | 0.32 | 0.39 | 0.9 | 0.4 | 0.39 | 0.43 | 0.33 | 0.38 | - |
| 60 | 0.55 | - | 0.39 | 0.43 | - | - | - | - | 0.5 | - | - | 0.34 | - | 1 | 0.42 | 0.44 | 0.45 | 0.35 | 0.41 | - |
| 70 | 0.6 | - | 0.41 | 0.45 | - | - | - | - | 0.55 | - | - | - | - | - | 0.44 | 0.49 | 0.47 | 0.37 | 0.44 | - |
| 80 | - | - | 0.43 | 0.47 | - | - | - | - | 0.6 | - | - | - | - | - | 0.46 | 0.54 | 0.49 | 0.39 | 0.47 | - |
| 90 | - | - | 0.45 | 0.49 | - | - | - | - | 0.65 | - | - | - | - | - | 0.48 | 0.59 | 0.51 | 0.41 | 0.5 | - |
| 100 | - | - | 0.47 | 0.51 | - | - | - | - | 0.7 | - | - | - | - | - | 0.5 | 0.64 | 0.53 | 0.43 | 0.53 | - |
| 110 | - | - | - | - | - | - | - | - | - | - | - | - | - | - | 0.52 | 0.69 | - | 0.45 | 0.56 | - |
| 120 | - | - | - | - | - | - | - | - | - | - | - | - | - | - | 0.54 | 0.74 | - | 0.47 | 0.59 | - |
| 130 | - | - | - | - | - | - | - | - | - | - | - | - | - | - | 0.56 | 0.79 | - | 0.49 | 0.62 | - |

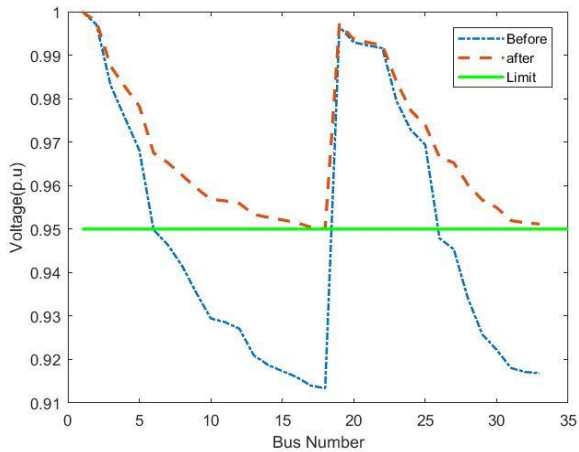

Fig. 2. Bus voltages before and after scenario A.

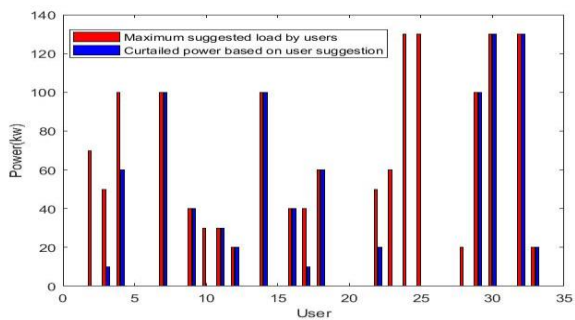

Fig. 3. Suggested load by consumers and their curtailed load by the market for A.

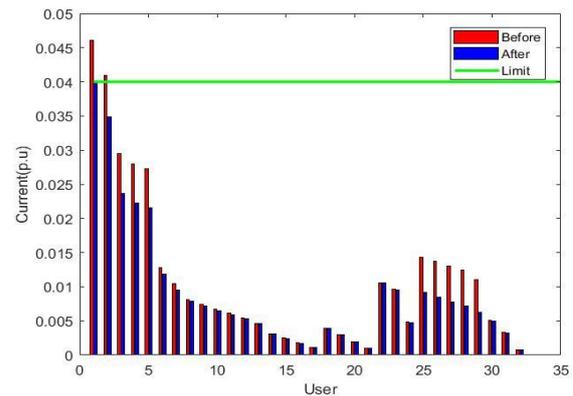

Fig. 4. Branch currents before and after B.

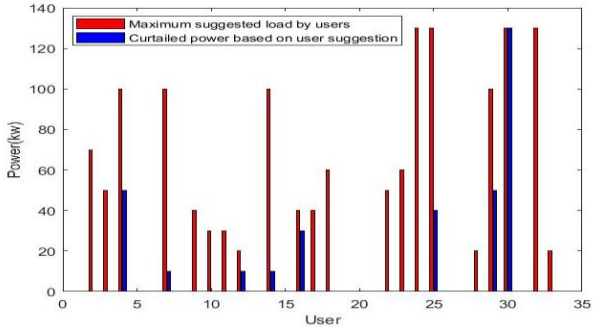

Fig. 5. Amount of curtailed load by the market for B and maximum proposed load by consumers

## C. Using market for Transmission System Operator (TSO) load reduction request

In this scenario, TSO aims to utilize LLRM to solve transmission system congestion. In this situation, feeder loss, branch congestion, and voltage drop are not essential issues for TSO, and it merely tries to reduce demand with the least cost. Consequently, constraints (6) and (7) can be neglected. Let us consider TSO wants to reduce 500 kW of feeder load, using LLRM the corresponding cost is 229.3$.

Table. 3. has summarized three discussed scenarios and presents to what level using LLRM can reduce the utility and TSO's costs.

TABLE III. COST OF SCENARIOS

| Scenario | Cost with LLRM ($) | Cost without LLRM ($) |
|---|---|---|
| A | 454.66 | 41328 |
| B | 141.7 | 15560 |
| C | 229.3 | 24369 |

## VI. CONCLUSION

In this paper, a Localized Load Reduction Market has been proposed using which power system entities like the utility company, TSO, and retailer can request for load reduction with various aims and consumers can actively participate in the market and provide bids for load reduction. After model formulation, the paper has presented GA for market clearance. Then, the application of this market has been studied and illustrated that in the situation of need for load reduction, the utility and TSO can cope with the problem at a very lower cost by running LLRM in comparison with conventional load curtailment.